\documentclass[aps,twocolumn,prb]{revtex4}
\usepackage[intlimits]{amsmath}
\usepackage{amssymb,amsfonts,graphicx}

\usepackage{defs-private}

\begin{document}

\title{Deterministic generation of N00N states using quantum dots in a cavity}

\author{Michael N. Leuenberger}\email{michael.leuenberger@ucf.edu}
\author{Mikhail Erementchouk}
\affiliation{NanoScience Technology Center and Department of Physics, University of Central
Florida, Orlando, FL 32826}

\begin{abstract}
\textbf{
Compared to classical light sources, quantum sources based on N00N states consisting of $N$ photons achieve an $N$-times higher phase sensitivity, giving rise to super-resolution.\cite{Giovannetti2004,Dowling,Giovannetti2011}
N00N-state creation schemes based on linear optics and projective measurements only have a success probability $p$ that decreases exponentially with $N$,\cite{Kok,VanMeter,Lee} e.g. $p=4.4\times 10^{-14}$ for $N=20$.\cite{Fiurasek} Feed-forward improves the scaling but $N$ fluctuates
nondeterministically in each attempt.\cite{Cable,McCusker}
Schemes based on parametric down-conversion suffer from low production efficiency and low fidelity.\cite{McCusker}
A recent scheme based on atoms in a cavity combines deterministic time evolution, local unitary operations, and projective measurements.\cite{Nikoghosyan}
Here we propose a novel scheme based on the off-resonant interaction of $N$ photons with four semiconductor quantum dots (QDs) in a cavity to create N00N states deterministically with $p=1$ and fidelity above 90\% for $N\lesssim 60$, without the need of any projective measurement or local unitary operation. Using our measure we obtain maximum $N$-photon entanglement $E_N=1$ for arbitrary $N$.
Our method paves the way to the miniaturization of N00N-state sources to the nanoscale regime, with the possibility to integrate them on a computer chip based on semiconductor materials.
}
\end{abstract}

\maketitle

In quantum metrology, the creation of a photonic Greenberger-Horne-Zeilinger (GHZ) state,\cite{GHZ} also known as N00N state\cite{Dowling} consisting of $N$ photons, 
can be used to achieve ultrahigh phase sensitivity,\cite{Giovannetti2004,Dowling,Giovannetti2011} providing viable mechanisms to beat the shot-noise
limit in optical interferometry and enabling super-resolution by beating the Rayleigh diffraction limit
by a factor of $N$.\cite{Bollinger,Boto,Kapale,Cable}
Since the demonstration that two-photon N00N states exhibit super sensitivity\cite{Kuzmich} and
super resolution\cite{Dangelo,Fonseca}, impressive progress has been made to experimentally realize
$N=3$,\cite{Mitchell} $N=4$,\cite{Bouwmeester,Walther,Nagata,Prevedel,Wieczorek,Matthews} and $N=6$ N00N states.\cite{Lu} 
Recently an $N=8$ N00N polarization state with a fidelity of 70\% has been created.\cite{Yao}
It is well known that coherent Schr\"odinger cat states 
$\ket{\psi_{CSC}}=\left(\ket{\alpha}+\ket{-\alpha}\right)/\sqrt{2+2e^{-2|\alpha|^2}}$ 
can be produced by a combination of Mach-Zehnder interferometer, cross-Kerr interaction, and postselection,\cite{Gerry}
which can also be used to create N00N states.\cite{GerryCampos:2001,GerryCampos:2002,Kapale}
Alternatively, N00N states can be produced by means of linear optics and postselection only,\cite{Kok,VanMeter,Lee,Cable}
with a success rate that decays exponentially with $N$.\cite{Fiurasek}
Here we demonstrate that $N$ photons in a cavity interacting off-resonantly with four QDs evolve deterministically from 
a non-entangled product state $\ket{N,0}$ into a N00N state $\ket{N::0}$ with probability $p=1$ for arbitrary $N$
and with fidelity above 90\% for $N\lesssim 60$.
The time evolution is governed by an effective photon Hamiltonian which we derive using our many-photon entanglement formalism\cite{Erementchouk:2010} based on Schwinger angular momentum operators.\cite{Yurke,Campos:1989}
The loading of the Fock state $\ket{N,0}$ into the cavity
can be done deterministically using stimulated Raman adiabatic passage (STIRAP).\cite{Brown} 
We adapt the loading process to QDs.

Our goal is to create the $N$-photon N00N state 
\be
\ket{N::0}=\frac{1}{\sqrt{2}}\left(\ket{N}_{+}\ket{0}_-+\ket{0}_+\ket{N}_{-}\right)
\label{eq:N00N_state}
\ee
in terms of the polarization degree of freedom, where $+$ and $-$ denote right and left circular polarization, resp.
Operating with a phase shifter for the $-$ mode of the form $U_{PG}=\exp(i\phi a_-\D a_-)$ on the N00N state, one obtains
\be
U_{PG}\ket{N::0}=\frac{1}{\sqrt{2}}\left(\ket{N,0}+e^{iN\phi}\ket{0,N}\right),
\label{eq:N00N_state_phase_shifted}
\ee
which improves the phase sensitivity by a factor of $N$ compared to a single-photon state 
\be
U_{PG}\ket{1::0}=\frac{1}{\sqrt{2}}\left(\ket{1,0}+e^{i\phi}\ket{0,1}\right).
\label{eq:single_photon_state_phase_shifted}
\ee

\begin{figure}[h]
  \includegraphics[width=3.3in]{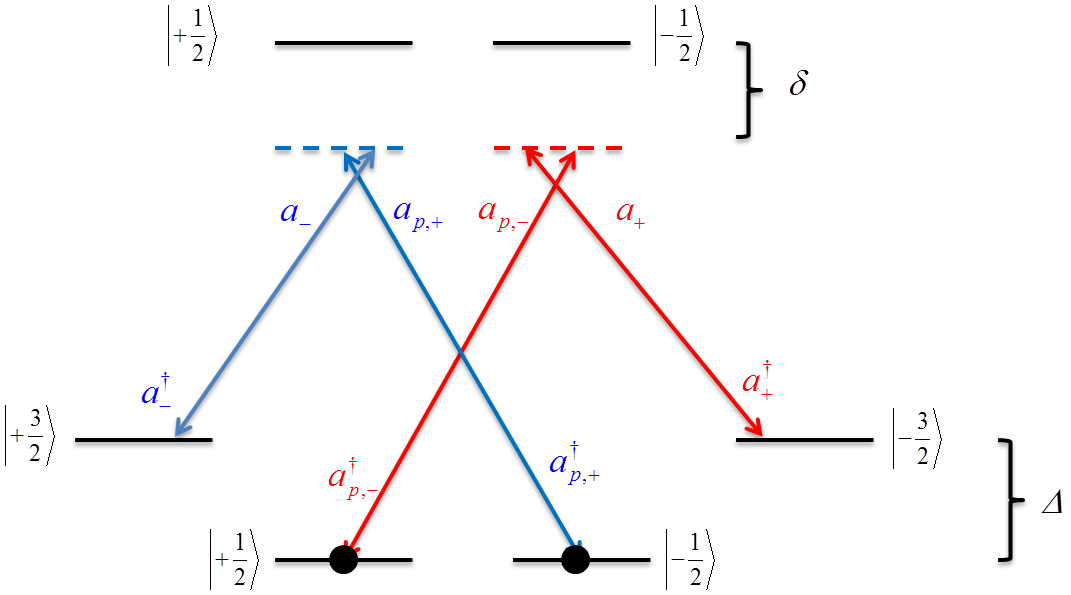}\\
  \caption{Quantum dot states that participate in the STIRAP process. The blue transitions give rise to
the $\ket{0,N}$ state, and the red transitions give rise to
the $\ket{N,0}$ state inside the cavity.
  }\label{fig:STIRAP_QD_levels}
\end{figure}

Let us first focus briefly on the loading of the Fock state $\ket{N,0}$ into the cavity.
It has been shown in Ref.~\onlinecite{Brown} that $N$ atoms can be used to create a $N$-photon Fock state
inside the cavity by means of STIRAP.
Instead of using atoms, we propose to use QDs with compressive strain in growth plane, such as standard GaAs QDs
embedded in Al$_x$Ga$_{1-x}$As. In the case of compressive strain, the light hole states lie below
the heavy hole states at an energy $\Delta$, as shown in Fig.~\ref{fig:STIRAP_QD_levels}.
The advantage of using QDs is that it is possible to select the three states connected by the red (blue) transitions
for creating the right-(left)-circularly polarized $\ket{N,0}$ ($\ket{0,N}$) state by means of a left-(right-)circularly polarized electric field of the pump laser
$\bE_{p,-}=\partial \bA_{p,-}/\partial t$ ($\bE_{p,+}=\partial \bA_{p,+}/\partial t$).
The classical pump laser field is given by the vector potential
\begin{equation}\label{eq:field_classical}
  \mathbf{A}_{p,\pm} = \frac{1}{(2\pi)^{3/2}}\int d^3k {\boldsymbol{\epsilon}}_{\bk,\pm} \frac{1}{\sqrt{2 \omega_{\bk,\pm}}}
  e^{i \mathbf{k}\cdot\bx} a_{\bk,\pm} + \text{h.c.},
\end{equation}
where $\boldsymbol{\mathbf{\epsilon}}_{\bk,\pm}$ are the unit polarization
vectors, $\bx$ is the coordinate, and
$a_{\bk,\pm}$ is the photon annihilation operator.

The Hamiltonian describing the STIRAP process in the rotating frame is given by
$H_{\rm STIRAP}=\delta\left(\ket{-\frac{1}{2}}_c\ket{0}\bra{-\frac{1}{2}}_c\bra{0}+\ket{\frac{1}{2}}_c\ket{0}\bra{\frac{1}{2}}_c\bra{0}\right)
+\Omega_{p,-}(t)[\ket{-\frac{1}{2}}_c\ket{0}\bra{\frac{1}{2}}_v\bra{0}+\ket{\frac{1}{2}}_v\ket{0}\bra{-\frac{1}{2}}_c\bra{0}]
+\Omega_{p,+}(t)[\ket{\frac{1}{2}}_c\ket{0}\bra{-\frac{1}{2}}_v\bra{0}+\ket{-\frac{1}{2}}_v\ket{0}\bra{\frac{1}{2}}_c\bra{0}]
+\Omega_{c,+}[\ket{-\frac{1}{2}}_c\ket{0}\bra{-\frac{3}{2}}_v\bra{1}+\ket{-\frac{3}{2}}_v\ket{1}\bra{-\frac{1}{2}}_c\bra{0}]
+\Omega_{c,-}[\ket{\frac{1}{2}}_c\ket{0}\bra{\frac{3}{2}}_v\bra{1}+\ket{\frac{3}{2}}_v\ket{1}\bra{\frac{1}{2}}_c\bra{0}]$
for each of the $N$ QDs.
The two dark states of each QD read
\be
\ket{\psi_{0,\pm}}  = C_\pm\left(\Omega_{c,\pm}\ket{\pm\frac{1}{2}}_v\ket{0}
+\Omega_{p,\mp}\ket{\mp\frac{3}{2}}_v\ket{1}\right)
\label{eq:dark_states}
\ee
where $C_\pm=1/\sqrt{\Omega_{p,\mp}^2+\Omega_{c,\pm}^2}$.
The eigenspace of zero eigenvalue for a particular number of $\pm$-polarized photons $l_\pm=n_{\mp 3/2,v}-q$ in the cavity is spanned by
\be
\ket{\psi_{0,q}^{N_\pm}}  = \frac{1}{Z_{q}^\pm}\sum_{j=q}^{N_\pm}
\frac{\left[-\Omega_{p,\mp}/\Omega_{c,\pm}\right]}{\sqrt{(N_\pm-j)!j!(j-q)!}}
\ket{N_\pm-j,0,j,j-q},
\label{eq:dark_states_N}
\ee
where $Z_{q}^\pm$ is a normalization constant and 
we adopted the second quantization state $\ket{n_{\pm 1/2,v},n_{\mp 1/2,c},n_{\mp 3/2,v},l_\pm}$. 
$n_{\pm 1/2,v}$, $n_{\mp 1/2,c}$, and $n_{\pm 3/2,v}$ count the number of QDs being in the state $\ket{\pm\frac{1}{2}}_v$, 
$\ket{\mp \frac{1}{2}}_c$, and $\ket{\mp\frac{3}{2}}_v$, resp.

\begin{figure}[h]
  \includegraphics[width=3.3in]{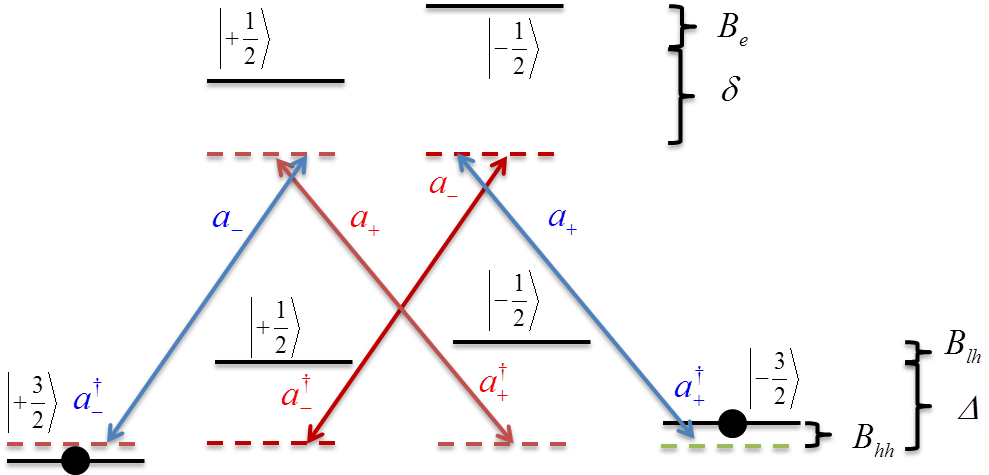}\\
  \caption{QD states of QDs 1 and 2 that participate in the N00N-state generation process. The blue (red) transitions
correspond to heavy- (light-) hole transitions.
  }\label{fig:N00N_QD1and2_levels}
\end{figure}

By adiabatically changing the pump field from $\Omega_{p,\mp}=0$ to $\Omega_{p,\mp}\gg \Omega_{c,\pm}$,
the coupled photon-QD system changes adiabatically from $\ket{N,0,0,0}$ to $\ket{0,0,N,N}$,
after which the pump field is switched off suddenly.

\begin{figure}[h]
  \includegraphics[width=3.3in]{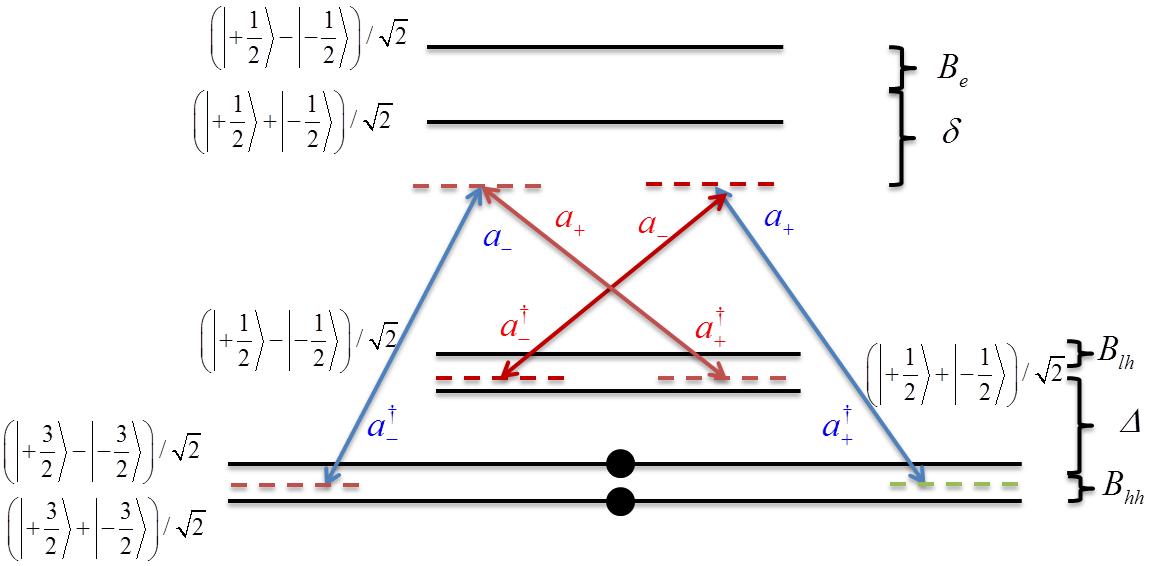}\\
  \caption{QD states of QDs 3 and 4 that participate in the N00N-state generation process. The blue (red) transitions
correspond to heavy- (light-) hole transitions.
  }\label{fig:N00N_QD3and4_levels}
\end{figure}
\begin{figure}[htb]
  \includegraphics[width=3.3in]{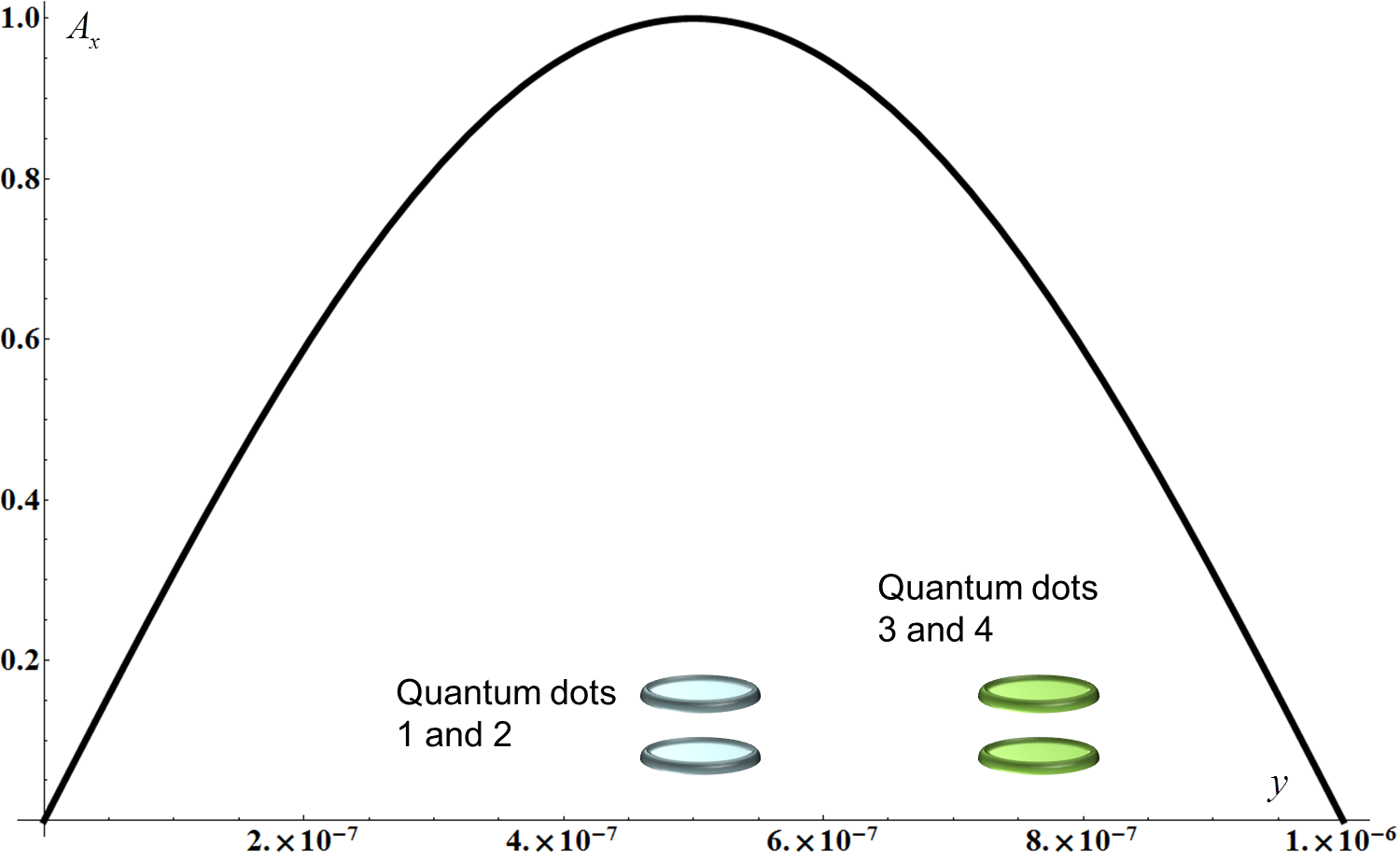}\\
  \caption{Minimum energy mode of the vector potential $A_x(y)$ inside a 3-dimensional rectangular cavity. In order to satisfy the condition $D_X+D_{XX}\ll A_X+A_{XX}$,
QDs 1 and 2 are located at large values of $A_x(y)$, whereas QDs 3 and 4 are positioned at smaller values of $A_x(y)$.
  }\label{fig:cavity_QDs}
\end{figure}

We describe now the minimal configuration we found that is able to create a N00N state of arbitrary photon number $N$.
Without loss of generality, we choose a left-circularly polarized pump field,
resulting in a $\ket{N_{+},0}$ Fock state, which can then be transferred
to an adjacent cavity containing four QDs that will
transform the $\ket{N_{+},0}$ Fock state to a $\ket{N::0}$ N00N state (see Fig.~\ref{fig:cavity_QDs}).
We choose the four QDs to have tensile strain, such that the heavy-hole states lie below the light-hole states, such
as already demonstrated for GaAsN QDs embedded in InP.\cite{Pohjola}
Figs.~\ref{fig:N00N_QD1and2_levels} and \ref{fig:N00N_QD3and4_levels} show the QD states of QDs 1, 2
and QDs 3,4, resp., that participate in the N00N-state generation process.
For our method to work, we need four magnetic fields $\bB_i$ that are applied individually to each QD $i=1,2,3,4$.
This can be achieved by means of local electric fields using the spin-orbit couplings to generate localized effective magnetic fields $\bB_i$.\cite{Nowack}
$\bB_1$ ($\bB_2$) points in $z$ ($-z$)-direction,
giving rise to the Zeeman splitting shown in Fig.~\ref{fig:N00N_QD1and2_levels} for QDs 1 and 2.
$\bB_3$ ($\bB_4$) points in $x$ ($-x$)-direction,
giving rise to the Zeeman splitting shown in Fig.~\ref{fig:N00N_QD3and4_levels} for QDs 3 and 4.

\textit{The key idea for creating the $\ket{N::0}$ N00N state is to use the off-resonant interaction of $\ket{N_{+},0}$
with the four QDs to generate a self-interaction between the $N$ photons,}
described by an effective $N$-photon Hamiltonian that we derive
by going up to 4th order in perturbation theory in this interaction $V$. The energy levels of the four QDs
are given by the unperturbed Hamiltonian
$H_0=\sum_{i=1}^2\{(\delta_i+B_{ei})\ket{-\frac{1}{2}}_{ci}\bra{-\frac{1}{2}}_{ci}+
(\delta_i-B_{ei})\ket{\frac{1}{2}}_{ci}\bra{\frac{1}{2}}_{ci}
+B_{hhi}\ket{-\frac{3}{2}}_{vi}\bra{-\frac{3}{2}}_{vi}+
B_{hhi}\ket{\frac{3}{2}}_{vi}\bra{\frac{3}{2}}_{vi}
+(\Delta+B_{lhi})\ket{-\frac{1}{2}}_{vi}\bra{-\frac{1}{2}}_{vi}+
(\Delta-B_{lhi})\ket{\frac{1}{2}}_{vi}\bra{\frac{1}{2}}_{vi}\}
\otimes\ket{N}\bra{N}
+\sum_{i=3}^4\{(\delta_i+B_{ei})\ket{s_-}_{ci}\bra{s_-}_{ci}+
(\delta_i-B_{ei})\ket{s_+}_{ci}\bra{s_+}_{ci}
+B_{hhi}\ket{s_{hh+}}_{vi}\bra{s_{hh-}}_{vi}+
B_{hhi}\ket{s_{hh+}}_{vi}\bra{s_{hh+}}_{vi}
+(\Delta+B_{lhi})\ket{s_{lh-}}_{vi}\bra{s_{lh-}}_{vi}+
(\Delta-B_{lhi})\ket{s_{lh+}}_{vi}\bra{s_{lh+}}_{vi}\}
\otimes\ket{N}\bra{N}$,
where $\delta_i$ are detuning energies of the cavity mode with respect to the QD transitions, 
the subscripts $_{ci}$ and $_{vi}$ denote the conduction and valence band states of QD $i$, 
$\ket{s_\pm}_{ci}=(\ket{\frac{1}{2}}_{ci}\pm\ket{-\frac{1}{2}}_{ci})/\sqrt{2}$,
$\ket{s_{hh\pm}}_{vi}=(\ket{\frac{3}{2}}_{vi}\pm\ket{-\frac{3}{2}}_{vi})/\sqrt{2}$,
$\ket{s_{lh\pm}}_{vi}=(\ket{\frac{1}{2}}_{vi}\pm\ket{-\frac{1}{2}}_{vi})/\sqrt{2}$.
The interaction of the $N$ photons with the four QDs is determined by
$V=\sum_{i=1}^2\{g_{lhi}(\ket{-\frac{1}{2}}_{ci}\bra{\frac{1}{2}}_{vi}
+\ket{\frac{1}{2}}_{ci}\bra{-\frac{1}{2}}_{vi})
+g_{hhi}(\ket{-\frac{1}{2}}_{ci}\bra{-\frac{3}{2}}_{vi}
+\ket{\frac{1}{2}}_{ci}\bra{\frac{3}{2}}_{vi})
\otimes\ket{N-1}\bra{N}\}
+\sum_{i=3}^4\{g_{lhi}(\ket{-\frac{1}{2}}_{ci}\bra{\frac{1}{2}}_{vi}
+\ket{\frac{1}{2}}_{ci}\bra{-\frac{1}{2}}_{vi})
+g_{hhi}(\ket{-\frac{1}{2}}_{ci}\bra{-\frac{3}{2}}_{vi}
+\ket{\frac{1}{2}}_{ci}\bra{\frac{3}{2}}_{vi})
\otimes\ket{N-1}\bra{N}\}+\mathrm{h.c.}
$,
where $g_{hhi}=\sqrt{N}A(\bx_i)P_{hhi}$, $g_{lhi}=\sqrt{N}A(\bx_i)P_{lhi}$ are the heavy-hole and light-hole constants
for the coupling between the $N$ cavity photons and QD $i$.
$A(\bx_i)$ denotes the photon field amplitude at the position $\bx_i$ of QD $i$,
and $P_{hhi}$ and $P_{lhi}=P_{hhi}/\sqrt{3}$ are the Kane interband matrix elements between the conduction band and the heavy- and light-hole band states, resp.
Since in general the Zeeman splittings differ among the bands due to the variation in effective mass, we use
$B_{e}$, $B_{hh}$, and $B_{lh}$ to denote the Zeeman splittings of the conduction, heavy-hole valence, and light-hole valence band states.

We choose the special symmetry $g_{hh1}=g_{hh2}$, $g_{hh3}=g_{hh4}$, $B_{1e}=-B_{2e}$, $B_{1hh}=-B_{2hh}$, $B_{1lh}=-B_{2lh}$, $B_{3e}=B_{4e}$, $B_{3hh}=-B_{4hh}$, and $B_{3lh}=-B_{4lh}$.
This combination ensures that all the linear $J_z$ terms vanish, so that there is no $J_z$ term that could lift the degeneracy between the $\ket{M=J}$ and $\ket{M=-J}$ states (see below).

\begin{figure}[h]
  \includegraphics[width=3.3in]{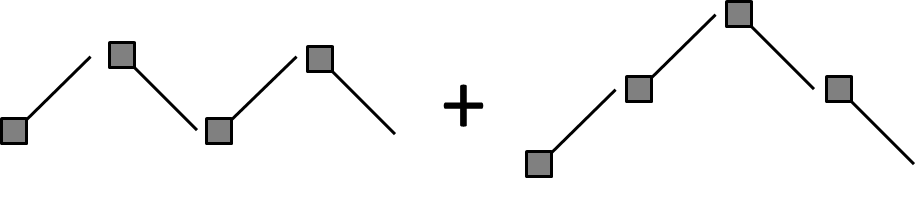}\\
  \caption{4th-order diagrams in $V$. The left term corresponds to two serially excited virtual excitons,
the right term corresponds to an excited virtual biexciton.
  }\label{fig:diagrams}
\end{figure}

We first switch to second-quantization operators for the photons by means of
$a_\pm=\sum_{n_\pm=0}^\infty\sqrt{n_\pm}\ket{n_\pm-1}\bra{n_\pm}$
and $a_\pm\D=\sum_{n_\pm=0}^\infty\sqrt{n_\pm}\ket{n_\pm}\bra{n_\pm-1}$.
Then we transform to the Schwinger representation
using the angular momentum operators 
$J_z=\frac{1}{2}(a_+\D a_+ - a_-\D a_-)$,
$J_+=a_+\D a_-$, $J_-=a_-\D a_+$, where
$J_\pm=J_x\pm iJ_y$.
Taking advantage of the off-resonant interaction, we make the approximation
that the excited states are not populated.
Consequently, all the odd powers in $V$ are approximately zero in the perturbation series.
Because of the special symmetry of the magnetic fields, 
all the 2nd-order terms in $V$, which correspond to the excitation of virtual single excitons, vanish as well.
This means the first nonvanishing contribution comes from 4th-order terms,
corresponding to the serial excitation of two non-interacting virtual excitons (denoted by subscript $_X$) 
and the excitation of virtual biexcitons (denoted by subscript $_{XX}$),
as depicted in Fig.~\ref{fig:diagrams}.\cite{Erementchouk:2010}
After long but straightforward calculations we arrive at the effective $N$-photon Hamiltonian
\be
H_N =A_N J_z^2+D_N \left(J_x^2-J_y^2\right)+B_NJ_x,
\label{eq:Hamiltonian_N}
\ee
where $n=n_++n_-=a_+\D a_+ + a_-\D a_-$, $A_N=\left(A_X+A_{XX}\right)$, $B_N=\left(B_{XX}+B_{XX}^{(n)}n\right)$, $D_N=\left(D_X+D_{XX}\right)$.
Note that the linear $J_x$ term comes from a 4th-order term
of the form $i[J_y,J_z]$.
$A_N$, $D_N$, and $B_N$ correspond to the longitudinal anisotropy, the transverse anisotropy and the effective magnetic field for the angular momentum $\bJ$, resp.
They are shown in the supplemental information.

In order to maximize entanglement, we get rid of the $B_N J_x$ term by choosing $\delta_3$ such that the numerator vanishes. From the resulting 3rd-order polynomial equation, we choose the real root for $\delta_3$.

\begin{figure*}[htb]
  \includegraphics[width=7in]{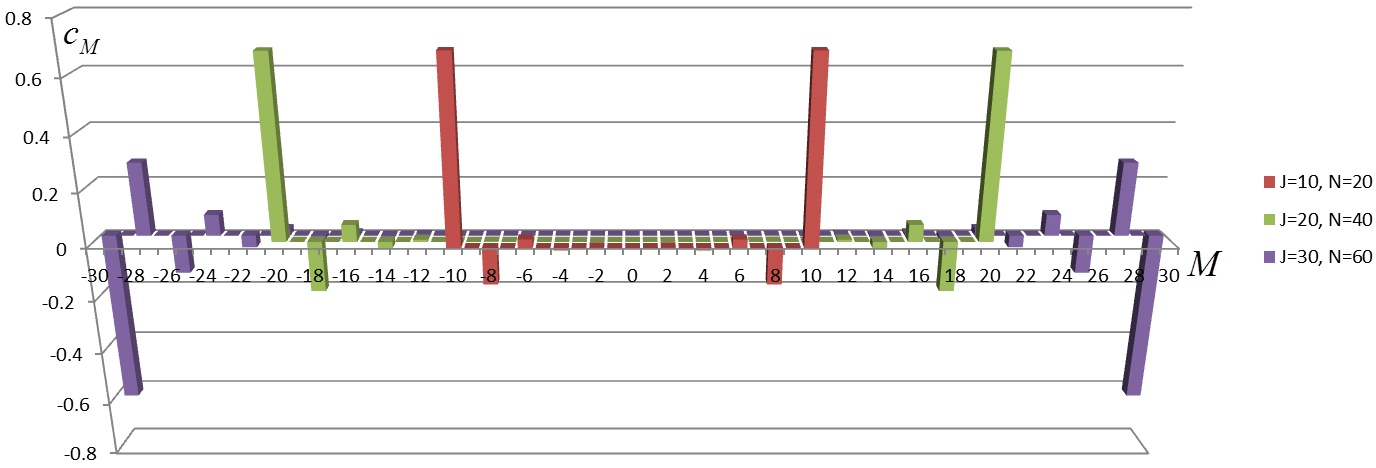}\\
  \caption{This histogram shows the amplitudes $c_M$ of the $N$-photon states $\ket{\psi(t=T_N/2)}$ after half a Rabi period $T_N/2$ for $N=20,40,60$,
corresponding to a total angular momentum of $J=10,20,30$, resp.
  }\label{fig:fidelity}
\end{figure*}

We discuss now our method to create the $\ket{N::0}$ N00N state.
After loading the cavity containing our four QDs with the Fock state $\ket{N_{+},0}$,
we let the system of $N$ photons evolve in time according to the Hamiltonian $H_N$ given in Eq.~(\ref{eq:Hamiltonian_N}).
The time evolution is best described in terms of the total angular momentum states of the $N$ photons.
Using the standard Schwinger representation
$\ket{J,M}=1/(\sqrt{(J+M)!(J-M)!})(a_+\D)^{J+M}(a_-\D)^{J-M}\ket{0}$ with $J=(N_++N_-)/2$, $M=(N_+-N_-)/2$
the time evolution corresponds to a Rabi oscillation between the states
$\ket{J,M=+J}=\ket{N_{+},0}$ and $\ket{J,M=-J}=\ket{0,N_{-}}$ in good approximation 
if $D_N\ll A_N$. 
At half the Rabi oscillation period $T_N/2$, we obtain our desired N00N state
\be
\ket{N::0}=\frac{1}{\sqrt{2}}\left(\ket{J,M=+J}+\ket{J,M=-J}\right).
\ee
This is the main result of this paper.
In order to satisfy the condition $D_N\ll A_N$ and stay in the regime of perturbation theory,
we must be able to choose $P_{hh1}$ independent of $P_{hh3}$.
This can be achieved by placing the QDs 1 and 2 at a position in the cavity that
has a different electromagnetic field amplitude than for the position of the QDs 3 and 4,
as shown in Fig.~\ref{fig:cavity_QDs}.

It has been shown in Refs.~\onlinecite{Agarwal,Zheng,Kapale} that an effective Hamiltonian of the form
$H_{\rm Rydberg}=\eta\left(-J_x^2+J_x\right)$ can be used to produce N00N states of atomic states of Rydberg atoms.
However, the time evolutions of $\ket{M=J}$ governed by $H_N$ and $H_{\rm Rydberg}$ are completely different.
While $H_{\rm Rydberg}$ leads the atomic state through all the $\ket{M}$ states with $M=-J,\ldots,J$ and therefore requires
very high phase accuracy and timing control to create the N00N state,
our effective Hamiltonian $H_N$ mixes in the regime $D_N\ll A_N$ only the states $\ket{M=J}$ and $\ket{M=-J}$,
which makes our method much more robust against phase and timing errors.

We present now our numerical results. Keeping in mind that photons can leak out of a cavity on timescales of $\tau_{md}=0.4$ $\mu$s for microdisk cavities\cite{Armani} with quality factor $Q=10^8$ and $\tau_{ms}=33$ ms for glass microsphere cavities\cite{Vernooy} with $Q=8\times 10^9$, we try to minimize the time it takes to create a N00N state.
Therefore we relax the condition $D_N\ll A_N$ to $D_N\lesssim A_N$. Consequently, we need to introduce
the fidelity $F_N=\left|\left<N::0|\psi(t=T_N/2)\right>\right|^2$ to describe the deviation of our result from the ideal N00N state.
In Fig.~\ref{fig:fidelity} we plot the amplitudes $c_M$ of the $N$-photon states $\ket{\psi(t=T_N/2)}=\sum_{M=-J}^J c_M\ket{M}$ for $N=20,40,60$, corresponding to a total angular momentum of $J=10,20,30$, resp.
For the calculations we use values that enhance the fidelity of the output states, while being realistic. 
The values common for all number of photons are 
$A(\bx_1)P_{hh1}=40$ $\mu$eV, 
$B_{e1}=-100$ $\mu$eV, $B_{e3}=500$ $\mu$eV, $B_{hh1}=-50$ $\mu$eV,
$B_{hh3}=460$ $\mu$eV, $B_{lh3}=460$ $\mu$eV, $\Delta=10$ meV,
$\delta_{1}=1000$ $\mu$eV.
Additionally for $N = 20$, $40$, $60$ we choose
$A(\bx_3)P_{hh3} = 2.5$, $3.0$ and $3.2$ $\mu$eV, resp.

For $N=20,40,60$ we obtain fidelities of $F_{20}=96\%$, $F_{40}=92\%$, $F_{60}=90\%$ and
half Rabi oscillation periods of $T_{20}/2=23$ ns, $T_{40}/2=0.67$ $\mu$s, $T_{60}/2=98$ $\mu$s.
Comparing to the lifetimes of the photons inside the cavities,
we see that a N00N state with $N=20$ photons can be already produced inside a microdisk cavity
due to $T_{20}/2\ll \tau_{md}$, and N00N states with $N=40$ and $N=60$ can be produced
inside a glass microsphere due to $T_{40}/2\ll \tau_{ms}$ and $T_{60}/2\ll \tau_{ms}$.

In order to calculate the many-photon entanglement, we use our many-particle expression for the concurrence\cite{Erementchouk:2010},
\be
C_N=\sqrt{4\mathrm{det}\aver{a_\kappa\D a_\lambda}/N}=\sqrt{1-\tilde{J}^2},
\ee
where $\tilde{J}=2|\aver{J}|/N$, $N=N_++N_-$, and $\kappa,\lambda=\pm$.
The entanglement of formation can then be calculated using Wootter's formula\cite{Wootters:1998}
$E_N=-\xi \log_2\xi-(1-\xi)\log_2(1-\xi)$, where $\xi=(1+\sqrt{1-C_N^2})/2$.
We obtain $C_N=1$ and $E_N=1$ for arbitrary $N$, i.e.
all the $N$-photon states $\ket{\psi(t=T/2)}$ are maximally entangled.
The reason is that the effective Hamiltonian $H_N$ in Eq.~(\ref{eq:Hamiltonian_N})
is quadratic in $\bJ$, because we made $B_N=0$.

Using $N$th-order perturbation theory, it is possible to approximate the Rabi oscillation energy by
$2D_N^{N/2}(A_NN^\eta)^{1-N/2}$, where $1/2<\eta<1$. Thus, in order to keep the fidelity high with increasing $N$,
we need $D_N/A_N=\lambda N^\eta$ with $0<\lambda\ll 1$ by choosing the appropriate locations
in the cavity for the QDs. This means $T_N/2$ increases exponentially as $1/(4D_N\lambda^{N/2-1})$.
This increase can be partially compensated by increasing $D_N$, which can be achieved by 
making the light-QD interaction larger and adding more QDs.

In conclusion, we have shown that N00N states with $N=20$, 40, and 60 photons are already
within experimental reach. The next challenge is to find a scheme based on light-QD interaction that
scales more favorably for the Rabi oscillation period.
Nevertheless, we developed here a scheme that provides the possibility for a 10-fold improvement over current state-of-the-art.
Our proposal paves the way to deterministic nanoscale N00N-state sources on a chip
based on semiconductor structures.

\acknowledgments

We acknowledge support from NSF (Grant ECCS-0901784),
AFOSR (Grant FA9550-09-1-0450), and NSF (Grant ECCS-1128597).


\section{Supplemental information}
We give here the coefficients of the Hamiltonian $H_N$.
\bea
A_X  & = &  \{64g_{hh3}^2g_{lh3}^2(B_{e3}^2-B_{hh3}^2-\delta_3^2)(B_{e3}^4\Delta +\Delta (B_{hh3}^2 \nn\\
& & -\delta_3^2)^2-2B_{e3}^2(B_{hh3}^2\Delta -2B_{hh3}B_{lh3}\delta_3+\Delta \delta_3^2))\} \nn\\
& & /\{(B_{lh3}^2-\Delta^2)(B_{e3}+B_{hh3}-\delta_3)^2(B_{e3}-B_{hh3} \nn\\
& & +\delta_3)^2(-B_{e3}+B_{hh3}+\delta_3)^2(B_{e3}+B_{hh3}+\delta_3)^2\},
\eea
\bea
D_X & = & \{128B_{hh3}B_{lh3}g_{hh3}^2g_{lh3}^2\delta_3(-B_{e3}^4+(B_{hh3}^2-\delta_3^2)^2)\} \nn\\
& & /\{(B_{lh3}^2-\Delta^2)(B_{e3}+B_{hh3}-\delta_3)^2(B_{e3}-B_{hh3} \nn\\
& & +\delta_3)^2(-B_{e3}+B_{hh3}+\delta_3)^2(B_{e3}+B_{hh3}+\delta_3)^2\}
\eea
\bea
B_{XX} & = & - \{4{B_{e3}}g_{hh3}^4(B_{e3}^3 - 2B_{hh3}^3 + B_{hh3}^2{\delta _3} - \delta _3^3 \nn\\ 
& & +B_{e3}^2(2{B_{hh3}} + 3{\delta _3}) + {B_{e3}}( - 5B_{hh3}^2 + \delta _3^2))\} \nn\\
& & /\{(B_{e3}^2 - B_{hh3}^2)({B_{e3}} - {\delta _3})(B_{e3}^2 - B_{hh3}^2 - 2{B_{e3}}{\delta _3} \nn\\
& & +\delta _3^2)(B_{e3}^2 - B_{hh3}^2 + 2{B_{e3}}{\delta _3} + \delta _3^2)\}   
\eea
\bea
B_{XX}^{(n)} & = & \{4{B_{e3}}g_{hh3}^4(B_{e3}^3 - 2B_{hh3}^3 + B_{hh3}^2{\delta _3} - \delta _3^3 \nn\\
& & + B_{e3}^2(2{B_{hh3}} + 3{\delta _3}) + {B_{e3}}( - 5B_{hh3}^2 + \delta _3^2))\} \nn\\
& & /\{(B_{e3}^2 - B_{hh3}^2)({B_{e3}} - {\delta _3})(B_{e3}^4 + {{(B_{hh3}^2 - \delta _3^2)}^2} \nn\\ 
& & - 2B_{e3}^2(B_{hh3}^2 + \delta _3^2))\}
\eea
\begin{widetext}
${A_{XX}} =  - \frac{{g_{hh1}^4{\delta _1}(31B_{e1}^3 - 31B_{hh1}^3 + 19B_{hh1}^2{\delta _1} - 29{B_{hh1}}\delta _1^2 + 17\delta _1^3 + B_{e1}^2( - 93{B_{hh1}} + 19{\delta _1}) + {B_{e1}}(93B_{hh1}^2 - 38{B_{hh1}}{\delta _1} + 29\delta _1^2))}}{{2(2{B_{e1}} - 2{B_{hh1}} + {\delta _1}){{({B_{e1}} - {B_{hh1}} + {\delta _1})}^3}{{( - {B_{e1}} + {B_{hh1}} + {\delta _1})}^3}}} - D_{XX}$.

Interestingly, $A_{XX}$ is a function of $D_{XX}$, which is given by
\bea
{D_{XX}} & = &  (2g_{hh3}^4(7B_{e3}^{12} + B_{e3}^{11}(2{B_{hh3}} + 5{\delta _3}) + 8B_{hh3}^2\delta _3^2{(B_{hh3}^2 - \delta _3^2)^4} - B_{e3}^{10}(15B_{hh3}^2 + 37\delta _3^2) \nn\\
& &  - B_{e3}^9(10B_{hh3}^3 - 147B_{hh3}^2{\delta _3} + 8{B_{hh3}}\delta _3^2 + 19\delta _3^3) \nn\\
& &  - 2B_{e3}^8(5B_{hh3}^4 + 62B_{hh3}^2\delta _3^2 - 39\delta _3^4) - {B_{e3}}{(B_{hh3}^2 - \delta _3^2)^3}(2B_{hh3}^5 + 25B_{hh3}^4{\delta _3} - 2B_{hh3}^3\delta _3^2 - 10B_{hh3}^2\delta _3^3 + \delta _3^5) \nn\\
& &  + 2B_{e3}^7(10B_{hh3}^5 - 223B_{hh3}^4{\delta _3} + 8B_{hh3}^3\delta _3^2 + 98B_{hh3}^2\delta _3^3 + 6{B_{hh3}}\delta _3^4 + 13\delta _3^5) \nn\\ 
& & + B_{e3}^6(50B_{hh3}^6 + 490B_{hh3}^4\delta _3^2 - 202B_{hh3}^2\delta _3^4 - 82\delta _3^6) \nn\\
& & - 2B_{e3}^5(10B_{hh3}^7 - 203B_{hh3}^6{\delta _3} - 3B_{hh3}^4\delta _3^3 + 2B_{hh3}^3\delta _3^4 + 167B_{hh3}^2\delta _3^5 + 4{B_{hh3}}\delta _3^6 + 7\delta _3^7) \nn\\
& &  + B_{e3}^4( - 45B_{hh3}^8 - 452B_{hh3}^6\delta _3^2 - 6B_{hh3}^4\delta _3^4 + 332B_{hh3}^2\delta _3^6 + 43\delta _3^8) \nn\\
& & + B_{e3}^3(10B_{hh3}^9 - 87B_{hh3}^8{\delta _3} - 16B_{hh3}^7\delta _3^2 - 268B_{hh3}^6\delta _3^3 + 4B_{hh3}^5\delta _3^4 + 350B_{hh3}^4\delta _3^5 + 4B_{hh3}^2\delta _3^7 + 2{B_{hh3}}\delta _3^8 + \delta _3^9) \nn\\
& &  + B_{e3}^2(13B_{hh3}^{10} + 115B_{hh3}^8\delta _3^2 + 226B_{hh3}^6\delta _3^4 - 346B_{hh3}^4\delta _3^6 + B_{hh3}^2\delta _3^8 - 9\delta _3^{10}))) \nn\\
& & /((B_{e3}^2 - B_{hh3}^2)({B_{e3}} - {\delta _3}){({B_{e3}} - {B_{hh3}} - {\delta _3})^3}{({B_{e3}} + {B_{hh3}} - {\delta _3})^3}{({B_{e3}} - {B_{hh3}} + {\delta _3})^3}{({B_{e3}} + {B_{hh3}} + {\delta _3})^3})
\eea
\end{widetext}

\end{document}